 \documentclass[preprint]{aastex}

\slugcomment{Accepted for ApJ Letters, 24 March 2000 }

\shorttitle{Polarised sub-mm emission from Sgr~A${^\star}$}
\shortauthors{D. Aitken et al.}

\begin{document}


\title{Detection of polarized mm and sub-mm emission from Sgr~A$^{\star}$}


\author{D.K. Aitken\altaffilmark{1,4}, J. Greaves\altaffilmark{2}, Antonio Chrysostomou\altaffilmark{1,2}}
\author{W. Holland\altaffilmark{1}, J.H. Hough\altaffilmark{2}, D. Pierce-Price\altaffilmark{3}}
\author{J. Richer\altaffilmark{3}}

\altaffiltext{1}{Dept. of Physical Sciences, University of
Hertfordshire, Hatfield, Herts, AL10 9AB, U.K.}

\altaffiltext{2}{Joint Astronomy Centre, 660 N. A`oh\=ok\=u Place,
University Park, Hilo, HI 96720, U.S.A.}

\altaffiltext{3}{Cavendish Laboratory, Madingley Road, Cambridge,
CB3 0HE, U.K.}

\altaffiltext{4}{email: d.aitken@star.herts.ac.uk}


\begin{abstract}
We report the detection of linear polarization from Sgr~A$^{\star}$ at
750, 850, 1350 and 2000$\mu$m which confirms the contribution of
synchrotron radiation. From the lack of polarization at longer
wavelengths it appears to arise in the millimetre/sub-millimetre
excess.  There are large position angle changes between the millimetre
and sub-millimetre results and these are discussed in terms of a
polarized dust contribution in the sub-millimetre and various
synchrotron models.  In the model which best explains the data the
synchrotron radiation from the excess is self-absorbed in the
millimetre region and becomes optically thin in the sub-millimetre.
This implies that the excess arises in an extremely compact source
$\sim$ 2 Schwarzschild radii.

\end{abstract}




\section{Introduction}
Observations of stellar proper motions in the vicinity of
Sgr~A$^{\star}$ (Ghez et al. 1998), the non-thermal radio source at
the apparent centre of the Galaxy, show that its mass of $\sim 2.5
\times 10^{6}~$M$_{\odot}$ is highly compact on a scale $<0.01$pc,
reinforcing its claim to be a massive black-hole candidate.  The
spectrum of Sgr~A$^{\star}$ extends as a rough power law $\sim
\nu^{\alpha}$, where $\alpha$ lies between $\onequarter$ and
$\onethird$, from a low-frequency turn-over at a few GHz up to $\sim$
100GHz, (e.g. Mezger, Duschl and Zylka 1996 and references therein),
and above this frequency there is evidence of a mm to sub-mm excess
over this power law extending almost to the atmospheric cut-off near
1000 GHz (Serabyn et al. 1997, Falcke et al. 1998). The nature of the
mm/sub-mm excess has been discussed by Serabyn et al. (1997) and the
possibility that it is due to dust is effectively eliminated: at 1.3mm
its size is less than 1 arcsecond with an implied brightness
temperature in excess of 100 K, inconsistent with the shorter
wavelength data, and the implication is that there are two separate
synchrotron components.

The radio spectrum of Sgr~A$^{\star}$ has for some years been
modelled in various ways in terms of synchrotron radiation but
searches for linear polarization, the characteristic signature of
this mechanism, have only recently been reported. Bower et al.
(1999a,b) used the VLA at 4.8, 8.4, 22 and 43~GHz in its
spectropolarimetric mode and the BIMA array at 86~GHz finding
upper limits of $<$0.2\%  at the lower frequencies and $<$1\% at
86~GHz.  These small upper limits on linear polarization, given
the sensitivity of the observations to large rotation measures due
to Faraday rotation (RM $\sim 10^{7}$ rad m$^{-2}$ at 8.4 GHz), are
difficult to account for and this is discussed at length by Bower
et al. (1999a). Apart from this there has been only a marginal
detection at 800 \micron~ of $4.9 \pm 3.2\%$ at 129 $\pm
19^{\circ}$ (Flett and Murray 1991) and a report of circular
polarization at 4.8 GHz by Bower et al. (1999c).

In mm to sub-mm polarimetric imaging studies of the central 15 pc
of the Galaxy we have detected linear polarization in
Sgr~A$^{\star}$ at 750, 850, 1350 and 2000 \micron~ which we
present here.  Discussion of the field distribution in the
circum-nuclear disk (CND) and neighbouring molecular clouds as
revealed by these observations will be presented separately.

\section{Observations and Results}

The observations were made using the SCUBA camera (Holland et al. 1999) and
its polarimeter (Greaves et al. 1999) on the 15m James Clerk Maxwell
Telescope (JCMT) in Hawaii. SCUBA observes simultaneously with
arrays of 37 and 91
bolometers, at either 850 and 450 $\mu$m respectively, or 750 and 350 $\mu$m
with a change of filters. Polarimetric observations were made in imaging
mode at all four submillimetre wavelengths, and single-point polarimetry
towards Sgr~A* was also done at 1.35 and 2mm, using single bolometers and
supplementing the data with small grid maps. The arrays have fields of view
of 2.3 arcmin, and full-width half-maximum beam sizes range from 7--8$''$ at
the shortest wavelengths to 34$''$ at 2 mm (Table 1). 

Observations were made by chopping the secondary mirror at 7.8 Hz to remove
the mean sky level, and nodding between left and right beams at slower rates
to take out sky gradients. For polarimetry, a rotating half-wave plate and
fixed etched grid were used to modulate the signal seen by the detectors.  
Polarimetry at 750 and 850 $\mu$m resulted in a Nyquist-sampled image after
32s of integration, after which the waveplate was stepped by 22.5$^{\circ}$.
The single-pixel data were observed in a similar manner but with 8s of
integration per waveplate angle. Total integration times were 5--10 complete
waveplate cycles, or 15--85 minutes. The data were reduced using the SURF
(Jenness and Lightfoot 1998) and POLPACK (Starlink user note 223) software
packages;  for each pixel a sinusoidal modulation is
fitted to deduce the source percentage and direction ($p,\theta$) of the
polarization.

From observations of Saturn and Uranus, during the run and from archival
data, the instrumental polarization is found to be close to 1\% and
known to $\pm 0.1$--$0.2$~\% at a given wavelength and elevation.  At 
450 and 1350$\mu$m the instrumental polarizations are larger
(3.5 and 2.7\% respectively due to wind blind structure)
but still known to 0.2 to 0.8\% respectively.
The planets were assumed to be
unpolarized and a limited check was made on this for Saturn where sky
rotation can be used to separate instrumental from any planetary
polarization.  While we cannot exclude the possibility that Saturn
is slightly polarized this introduces an error of less than 0.3\% in
the instrumental polarization subtraction.  The instrumental polarization
changes by only 0.1\% over the elevation range used and the uncertainties
in the corrected polarization is considered good to $\pm 0.1-0.2$\%. 
Errors in the absolute orientation of the waveplate are one degree or less.

Intensity calibration at 1350 and 2000$\mu$m was obtained using Uranus
taking the brightness temperatures as 96K and 110K respectively,
and at 850$\mu$m from Mars taking $T_{B}$~=~208K. Since the
emphasis is on the polarimetric results, we did not perform extremely
accurate calibration observations, but typical uncertainties are only about
$\pm 10$~\%. For the 750 $\mu$m data Saturn was used for calibration taking
$T_{B}$=123K and a
greater error was introduced since this planet is larger than the beam; we
note that the 750 $\mu$m flux for Sgr~A$^{\star}$ appears anomalously low.
Atmospheric tranmission was measured with skydips.

A fundamental limitation is the maximum chop throw of 180$''$ (in this case
at a position angle of 145$^{\circ}$ east of north), which proved to be
insufficient at 350 and 450 $\mu$m. Surrounding dust emission has a steep
spectral index (Pierce-Price et al.. 2000) resulting in increasing off-beam
contamination at these shortest wavelengths where Sgr~A$^{\star}$
was not detected above background.
At 850 $\mu$m the off-beam
signals are only 15\% of the Sgr~A$^{\star}$ flux and these uncertainties
should be similar or smaller at 750, 1350 and 2000 $\mu$m at all of
which Sgr~A$^{\star}$ was clearly detected.
During the observation period the field of view rotates but the chop
orientation on the sky is maintained; thus every
spatial point in the map retains the same chop positions in $\alpha$
and $\delta$ although different array bolometers may be involved.
The results of polarimetric imaging in the vicinity of Sgr~A$^{\star}$
are presented  in Table 1 and the total flux at the position of
Sgr~A$^{\star}$ is given in column 3.

There will in general be two contributions to the
background, namely from cool dust in the inner cavity and the edges
of the CND and free-free emission from the HII region Sgr~A West,
predominantly from the central ionized filaments.

Dust and free-free emission contribute significantly to the observed flux
but in this wavelength region there is no independent evidence
at sufficient resolution to determine the contribution of the dust emission
component in the central beams.   In the sub-mm the flux in annuli from one
to two beam radii were determined from the images and at 1350 and
2000$\mu$m observations in eight beams circumferentially distributed about
SgrA$^{\star}$ sampled the ambient flux between 1.7--3.7 beam radii from
SgrA$^{\star}$.
Estimates of the free-free contribution in the central beams and these
peripheral areas were made from the 2 arcsecond resolution
3.6cm radio continuum maps of Roberts and Goss (1993),
after removing the Sgr~A$^{\star}$ point source itself.
These fluxes were then scaled to the present wavelengths as $\nu^{-0.1}$,
appropriate for optically thin free-free emission, and the approximation
is made that the central dust contribution is just that derived from the
peripheral regions after subtraction of the free-free component 
(column 7 in the table).  The flux from SgrA$^{\star}$ after subtraction of
the dust and free-free contributions (column 8)
will also include any local dust excess (or deficit) over background.
Since at 2mm the correction
for free-free is large due to the large beam size the 
reliability of the non-thermal flux estimate at this wavelength will be
affected.

The observed E vector polarizations and position angles (north through
east) are listed in Table 1 in columns 9 and 10.  At 450$\mu$m there
is general polarization over the central 30 arcseconds with no
significant change at the position of Sgr~A$^{\star}$ and its average
$\sim$ 3\% at a position angle of 100$^{\circ}$ (E vector) is
attributed to dust.  At wavelengths between 750 and 2000$\mu$m the
polarization at the position of Sgr~A$^{\star}$ is well detected at
the 10-$\sigma$ level.  In the sub-mm it stands out from its
surroundings and is clearly associated with the flux peak at the
position of Sgr~A$^{\star}$.  This can be seen in Fig 1 which shows
the central 80 $\times$ 100$''$ region about Sgr~A$^{\star}$ at
850$\mu$m.  At 1350 and 2000$\mu$m the observations are from single
pointed observations where the HII region Sgr~A West and
Sgr~A$^{\star}$ are the dominant sources in the beam and
Sgr~A$^{\star}$ is the only significant contributor to polarization.

%% In this section, we use  the \subsection command to set off
%% a subsection.  \footnote is used to insert a footnote to the text.


%% This section also includes several of the displayed math environments
%% mentioned in the Author Guide.

\section{Interpretation}

At 2000$\mu$m, and to a lesser extent at 1350$\mu$m, there are only
two contributors to the flux: diffuse free-free emission and
Sgr~A$^{\star}$ itself.  Since the
former is unpolarized the polarization observed confirms the synchrotron
nature of Sgr~A$^{\star}$; it also suggests that the polarization is only
associated with the mm/sub-mm excess since it has not been observed at
wavelengths of 3.5mm and longer where the power-law spectrum dominates.
Note that at 2000$\mu$m the largest contributer is from free-free emission
and therefore the intrinsic polarization of SgrA$^{\star}$ must be large.

A large difference of position angle between the mm and
sub-mm observations is evident in Table 1 and various ways of
understanding this are next investigated.

Although the mm position angles themselves are very close the three
shortest wavelength position angles define a straight line when
plotted against $\lambda^{2}$; if a rotation of $\pi$ is subtracted
from the 2mm angle a rough linear relation is maintained.  However the
rotation measure then indicated is so large ($-1.44\times10^{6}$
rad/m$^{2}$ = $\pi/(\lambda_{2}^{2} - \lambda_{1.35}^{2})$) that
polarization within the band pass of 40GHz at 2mm would be
undetectable.  Therefore the closeness of the mm position angles
indeed indicate that there is little if any Faraday rotation.

Changes in polarization and position angle with time are well documented
in the compact cores of blazars and some AGN (e.g. Saikia and Salter
1988) and also have been observed at mm wavelengths (Nartallo et al. 1998).
Sgr~A$^{\star}$ is known to be variable on a time scale of
months.  While there is a  separation of some months between the
observations presented here, those at 750, 1350 and 2000$\mu$m
were all made in August 1999 within a week and these do show the large
position angle difference between the mm and sub-mm polarizations.
Furthermore, the 850$\mu$m observation in March also lies
on the same trend of position angle with wavelength.
Variability is therefore unlikely to have been a significant influence on
the position angle, although it seems possible that
Sgr~A$^{\star}$ may have increased in brightness during this period.

Since the sub-mm dust emission is polarized it will affect the
observed polarization and it is tempting to try to attribute the whole
position angle change to dust.

Denoting by $I_{o}, I_{d}$ and $I_{s}$ the observed central flux,
the estimated dust and Sgr~A$^{\star}$ components respectively and by
$q_{o}, q_{d}$ and $q_{s}$ their Stokes $q$ parameters, the
synchrotron component $q_{s}$ can be found from
$$  I_{o}q_{o} = I_{d}q_{d} + I_{s}q_{s}   $$
and similarly for the Stokes $u$ components.

It is difficult to estimate the dust contribution to polarization from
the 750 and 850$\mu$m observations because of the proximity of
Sgr~A$^{\star}$ and its high intrinsic polarization.  Instead we use
the 450$\mu$m observations in which Sgr~A$^{\star}$ is not detected,
and the polarization is $\sim$ 3\% at a position angle of $\sim
100^{\circ}$.  There are some problems here with polarized flux in the
reference beams but other indicators of dust alignment in the central
parsec are consistent with this estimate: observations at 350$\mu$m
(Novak et al. 2000) yield 1.9\% at 87$^{\circ}$ near the position of
SgrA$^{\star}$, and at 100$\mu$m the results are similar (Hildebrand
et al. 1993) though both these are in a larger beam; the only
information at high resolution is from the mid-infrared (Aitken et
al. 1996) where integrations over a central 14 arcsecond diameter
yield 2.4\% at 125$^{\circ}$, admittedly weighted to warm dust.

The intrinsic polarization resulting from removing the unpolarized
free-free contribution and adopting 3\% at 100$^{\circ}$ for dust
emission is shown in Table 1 columns 11 and 12.  At mm wavelengths
these numbers are insensitive to the uncertainties in the adopted dust
polarization and even in the submm the range of dust polarizations
given in the previous paragraph changes the intrinsic polarization by
$\pm$3\% at most and less than $\pm$3$^{\circ}$. Similarly
uncertainties in the relative fractions of the three flux components
has little effect on the intrinsic position angle but changing
dilution does introduce errors to the intrinsic polarization fraction
and at 750 and 2000$\mu$m the upper bounds are poorly defined. In
Table 1 the errors to the intrinsic polarization are derived by
assuming $\pm$15\% errors to the free-free and $\pm$20\% errors to the
dust contributions except at 2000$\mu$m where the adopted error is
$\pm$40\% (these dust contribution errors are derived from the
observed variation in the peripheral flux).  It is clear that dust
polarization cannot explain the abrupt change of position angle
between the mm and sub-mm, rather it is increased and it approaches
$\pi$/2.

We can turn the above argument around to determine what dust polarization
is needed in the sub-mm to cause the position angle shift given the
intrinsic synchrotron polarization of the mm results, say $\sim10$\%
at 85$^{\circ}$.  With the 850$\mu$m fluxes and polarization given
in Table 1 the required dust polarization is then 12.0\% at 163$^{\circ}$,
larger than any dust emission polarization yet observed.  It might be
argued that this could be due to unusual conditions close to Sgr~A$^{\star}$
and an additional dust component in the central beam.  Any such emission
is already included in the Table 1 entry for Sgr~A$^{\star}$ and because the
same polarized intensity ($\sim450$mJy) has to be supplied from  much
less dust emission the required polarization would be in excess of 25\%.

An assumption made so far is that throughout the wavelength region 
of these observations  the synchrotron radiation remains
either optically thin or self-absorbed.
Most models of the synchrotron SED in Sgr~A$^{\star}$ consider the
mm/sub-mm excess to be self absorbed but it may well be that at the
shortest wavelengths it becomes optically thin.  
In that  case the synchrotron position angle shifts through 90$^{\circ}$
and becomes close to 175$^{\circ}$ in the sub-mm.  The dust
corrected sub-mm points are now discrepant by 6 to 14$^{\circ}$ and 
this could be attributed to the crudeness of the correction.

Approximating the synchrotron emitting region by a simple slab with
constant energy density and magnetic field we find (e.g. Ginzburg and
Syrovatskii 1965, 1969, Pacholczyk and Swihart 1967)
that for the polarization to change sign near 1mm
then $\nu_{ssa}$, the synchrotron self absorption frequency, lies in
the range 0.6--0.7mm, close to the peak of the mm/sub-mm excess, and the
required fields, densities and energies are similar to those in the
compact source models of  Beckert and Duschl (1997).  The polarization
transition of this simple model is quite sharp, as is observed,
with the polarization
quickly approaching  saturation on the long wavelength side,
changing sign near 1mm and steadily rising at shorter wavelengths.

A variant on this explanation is that self-absorption results in different
optical depths with possibly different field distributions
being probed as a function of wavelength.   In that case, however,
one should encounter diminishing polarization towards shorter
wavelengths due to the superposition of a range of field orientations,
and this is not observed.

%% This section contains more display math examples, including unnumbered
%% equations (displaymath environment). The last paragraph includes some
%% examples of in-line math featuring a couple of the AASTeX symbol macros.

\section{Discussion}

%% The displaymath environment will produce the same sort of equation as
%% the equation environment, except that the equation will not be numbered
%% by LaTeX.

Synchrotron radiation from Sgr~A$^{\star}$ has been detected
between 750$\mu$m and 2000$\mu$m with an intrinsic polarization
$\sim$ 10\%.  This polarization arises from the spectral region of the
mm/sub-mm excess.  
The present observations are compounded by the presence of polarized
emission from dust and dilution by free-free emission in these large
beams.  Although this has led to a number of possible explanations for
the large position angle shift between the mm and sub-mm results all
but one of these can be eliminated:
the closeness of the position angles at 1.35 and 2mm require that there is
little or no Faraday rotation,
dust polarization can only produce a small shift which is in the wrong
sense, and the shift is not due to variability.
Only a transition between optically thin synchrotron radiation in
the sub-mm and self-absorption in the mm region appears as a plausible
explanation of the shift.
Future small beam observations will be needed to confirm
this result.

A puzzle re-emphasized here is the lack of polarization at 3.5mm
(Bower et al.. 1999a,b) and longer wavelengths in view of the large
intrinsic polarization of Sgr~A$^{\star}$  at 1.35 and 2mm.  Sufficient
dilution of the mm/sub-mm excess to bring it below 1\% requires  not only
a steeply falling spectrum with decreasing frequency for the
excess, where $\nu^{5/2}$ is expected for self absorption, but also that
the cut-off frequency of the power-law section is below 2mm to ensure a
fast rising flux of the power law region.
It is possible, as well, that a flare in Sgr~A$^{\star}$
at 100GHz in March 1998 (Tsuboi, Miyazaki and Tsutsumi 1998)
has affected the result of Bower et al. 1999b.

The synchrotron self-absorption transition in the sub-mm proposed here
is consistent with the model of the excess by Beckert and Duschl (1997),
and requires the excess to arise from a compact source $\sim$ 2
Schwarzschild radii in size.
Also because of the transition the $\pi$/2 ambiguity of the
relationship between position angle and magnetic field is removed
and hence the field orientation is close to east-west, since at mm
wavelengths the emission is considered self-absorbed.

\section{Conclusions}
We report the observation of millimetre and sub-millimetre
polarization from Sgr~A$^{\star}$, confirming the role of synchrotron
radiation. The polarization is a property of the mm/sub-mm excess,
demonstrating that the excess is real and not an artefact of
variability and that it and the power-law spectrum arise in distinct
structures. There is a large position angle shift between the mm
and sub-mm observations. This can be explained by a transition
between optically thin and self absorbed synchrotron radiation near
1~mm.  Such a high self-absorption frequency implies a very compact
source $\sim$ 2R$_{\rm S}$. 


%% Included in this acknowledgments section are examples of the
%% AASTeX hypertext markup commands. Use \url without the optional [HREF]
%% argument when you want to print the url directly in the text. Otherwise,
%% use either \url or \anchor, with the HREF as the first argument and the
%% text to be printed in the second.

\acknowledgments

We are grateful to the UK Panel for Allocation of Telescope Time for
the award of observing time for this project. JSR acknowledges a Royal
Society Fellowship and DP-P a PPARC Research Studentship. The JCMT is
operated by the Joint Astronomy Centre on behalf of PPARC of the UK,
the Netherlands OSR, and NRC Canada. The authors would like to thank
the NCSA Astronomy Digital Image Library (ADIL) for providing images
for this paper. We thank an unknown referee for useful comments.

\clearpage
\begin{figure}
  \centering
  \epsscale{0.95}
  \plotone{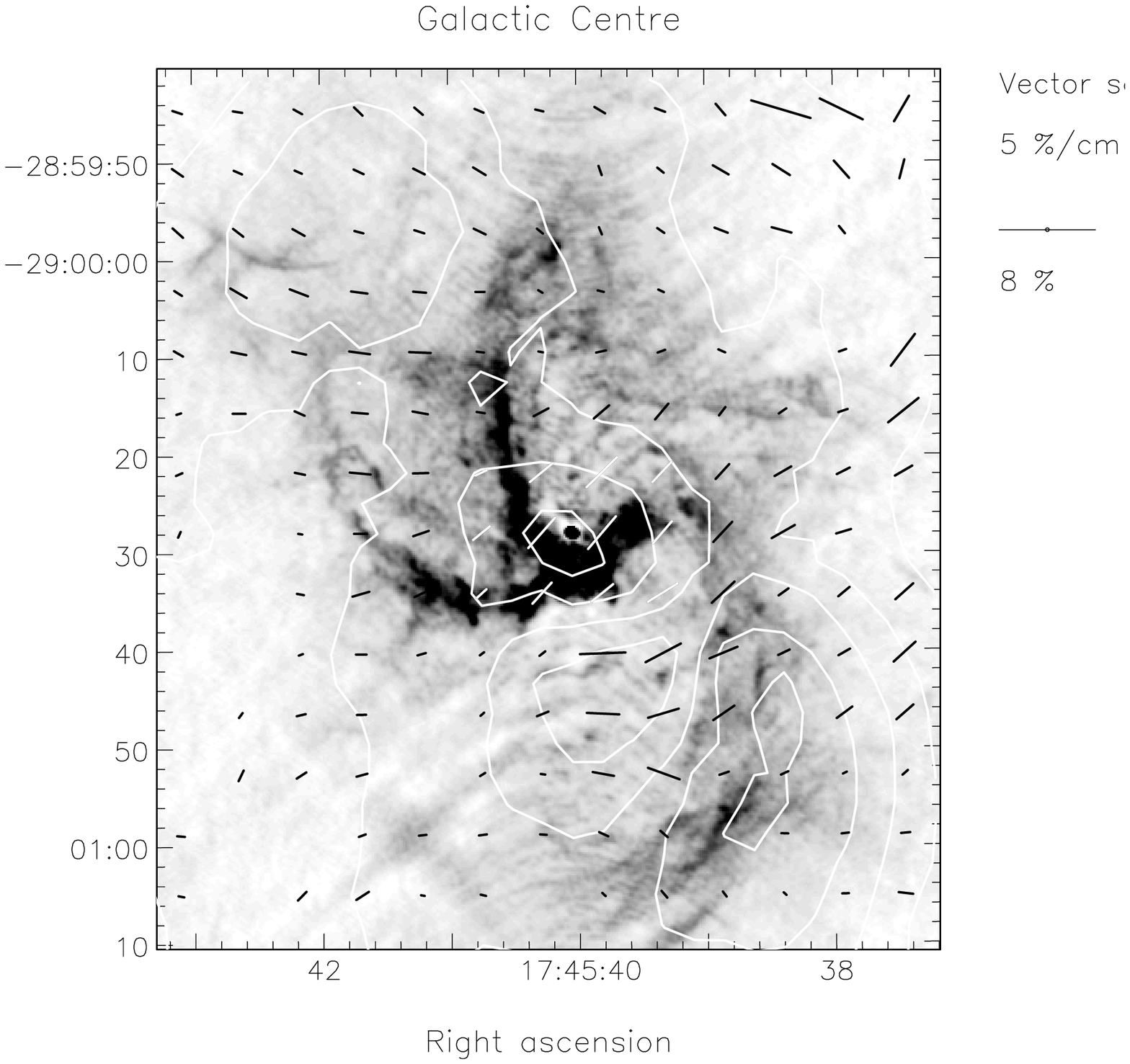}
\end{figure}

%% Generally speaking, only the figure captions, and not the figures
%% themselves, are included in electronic manuscript submissions.
%% Use \figcaption to format your figure captions. They should begin on a
%% new page.

\clearpage

%% No more than seven \figcaption commands are allowed per page,
%% so if you have more than seven captions, insert a \clearpage

%% There must be a \figcaption command for each legend. Key the text of the
%% legend and the optional \label in curly braces. If you wish, you may
%% include the name of the corresponding figure file in square brackets.
%% The label is for identification purposes only. It will not insert the
%% figures themselves into the document.
%% If you want to include your art in the paper, use \plotone.
%% Refer to the on-line documentation for details.

\figcaption[fig1.eps]{Greyscale image of 3.6 cm emission (Roberts \&
Goss 1993, courtesy of ADIL) overlaid with contours of the 850
\micron~ emission from the central region around Sgr~A$^{\star}$,
which is clearly seen as a point source in the centre of this
image. The short lines on the figure show the magnitude and direction
of the sub-millimetre polarisation {\bf E}-vector. Note how the
magnitude of the polarisation is larger at the position of
Sgr~A$^{\star}$ (more than one large vector is seen as the beam is
bigger than the vector spacing in this figure). Coordinates for
SgrA$^{\star}$ are 17$^{\rm h}$~45$^{\rm m}$~40.049$^{\rm s}$, -29\degr~00\arcmin~27.98\arcsec~(J2000).
\label{fig1} }

%% Tables should be submitted one per page, so put a \clearpage before

%% Two options are available to the author for producing tables:  the
%% deluxetable environment provided by the AASTeX package or the LaTeX
%% table environment.  Use of deluxetable is preferred.
%%

%% Three table samples follow, two marked up in the deluxetable environment,
%% one marked up as a LaTeX table.

%% In this first example, note that the \tabletypesize{}
%% command has been used to reduce the font size of the table.
%% Note also that the \label command needs to be placed

\clearpage

\begin{deluxetable}{cc|ccc|ccc|cccc|c}
\tabletypesize{\scriptsize}

\tablecaption{{\bf Sgr~A$^{\star}$ fluxes} }

\tablehead{ 

\colhead{$\lambda$} & \colhead{beam}   &
\multicolumn{3}{c}{central flux (Jy/beam)} & \colhead{ambient} &
\colhead{central} & \colhead{Sgr~A${^\star}$} &
\multicolumn{2}{c}{obs. polarsn.} &
\multicolumn{2}{c}{intr. polarsn.} & \colhead{date} \\

\colhead{\micron~} & \colhead{arcsec} & \colhead{total} &
\colhead{excess}  & \colhead{free-free}  & \colhead{free-free} &
\colhead{dust} & \colhead{ } & \colhead{P(\%)} & \colhead{$\theta
(^\circ)$ } & \colhead{P(\%)} & \colhead{$\theta (^\circ)$ } & \colhead{ } \\

\colhead{1} & \colhead{2} & \colhead{3} &
\colhead{4\tablenotemark{a}} & \colhead{5} & \colhead{6} &
\colhead{7\tablenotemark{b}} & \colhead{8} & \colhead{9} &
\colhead{10} & \colhead{11} & \colhead{12} & \colhead{13} \\

 \colhead{} & \colhead{} & \colhead{} & \colhead{} & \colhead{} &
\colhead{} & \colhead{=3-4-6} & \colhead{=4+6-5} & \colhead{} &
\colhead{} & \colhead{} & \colhead{} & \colhead{} 

}

\startdata
  750 &  12.5\tablenotemark{c} & $\sim$5 &$\sim$2.5 & 2.5 & 1 & $\sim$1.5& $\sim 1\pm0.5$ &  3.8$\pm$0.4   & 164$\pm$3 & 22{$^{+25}_{-9}$} & 169$\pm$3 & 27 Aug 99\\
  850 &  14 &  8.2 & 3.6 & 2.25 & 0.95 & 3.65 & 2.3$\pm$0.9 &  3.3$\pm$0.2  & 150$\pm$2 & 13{$^{+10}_{-4}$} & 161$\pm$3 & 25 Mar 99\\
  1350 & 22.5 & 6.5 & 4.9 & 3.55 & 0.85 & 0.75 &2.2$\pm$0.5 & 4.1$\pm$0.4 &89$\pm$3 & 11{$^{+3}_{-2}$} & 88$\pm$3  & 22 Aug 99\\
  2000 & 33.5 & 8.0 & 6.4 & 5.8 & 1.2 & 0.4 & 1.8$\pm$0.9 & 2.9$\pm$0.3 & 84$\pm$3 & 12{$^{+9}_{-4}$} & 83$\pm$3 & 22 Aug 99\\
\enddata

%% Text for table notes should follow after the \enddata but before
%% the \end{deluxetable}. Make sure there is at least one \tablenotemark
%% in the table for each \tablenotetext.

\tablenotetext{a}{(central excess)=(central total)-(ambient)}
\tablenotetext{b}{flux from the central dust component is taken as equal to the flux from the ambient dust}
\tablenotetext{c}{fluxes were measured in a 16 arcsec beam}

\tablecomments{Errors on observed P and $\theta$ are 1$\sigma$ statistical
errors. See text for uncertainty estimates on Sgr~A{$^\star$} flux
and intrinsic polarisations.}

%\tablecomments{Occasionally, authors wish to append a short paragraph
% of explanatory notes that pertain to the entire
% table, but which are different than the caption.  Such notes should be
% placed in a {\tt tablecomments} command like this.}

\end{deluxetable}

%% The following command ends your manuscript. LaTeX will ignore any text

\end{document}